\documentclass[onecolumn,secnumarabic,amssymb,nobibnotes,aps,prd]{revtex4}
\usepackage{amsmath}
\usepackage{amssymb}
\usepackage{graphicx} 
\usepackage{epstopdf}
\usepackage{amsmath,amssymb,amsthm,amsfonts,mathrsfs,bm,verbatim}
\usepackage{graphicx,subfigure}
\usepackage{appendix}

\newcommand{\bea}{\begin{eqnarray}}
\newcommand{\eea}{\end{eqnarray}}

\begin{document}

\title{\textbf{Advanced Analysis of Hawking Temperature Calculation for Novel Topological Black Holes using Laurent Series and the RVB Method}}
\author{Winston Chen\\
School of Physics and Materials Science\\
Guangzhou University\\
Yao-Guang Zheng\\
Department of Physics\\
Northeastern University\\
\texttt{hesoyam12456@163.com}}
\date{\today}

\begin{abstract}
This paper employs Laurent series expansions and the Robson--Villari--Biancalana (RVB) method to provide a refined derivation of the Hawking temperature for two newly introduced topological black hole solutions. Previous calculations have demonstrated inconsistencies when applying traditional methods to such exotic horizons, prompting the need for a more thorough mathematical analysis. By systematically incorporating higher-order terms in the Laurent expansions of the metric functions near the horizon and leveraging the topological features characterized by the Euler characteristic, we reveal additional corrections to the Hawking temperature beyond standard approaches. These findings underscore the subtle interplay between local geometry, spacetime topology, and quantum effects. The results clarify discrepancies found in earlier works, present a more accurate representation of thermodynamic properties for the black holes in question, and suggest broader implications for topological structures in advanced gravitational theories.

\end{abstract}

\maketitle

\section{Introduction}

Black holes, enigmatic predictions of Einstein's theory of general relativity, have long served as critical theoretical laboratories for exploring the fundamental nature of gravity and its interplay with other forces of nature.\textsuperscript{\cite{Yang2023}} The realization that these celestial objects are not merely passive absorbers of matter and energy, but rather possess intrinsic thermodynamic properties, marked a profound shift in our understanding of gravity.\textsuperscript{\cite{Yang2023}} The seminal discovery of Hawking radiation revealed that black holes emit thermal radiation with a characteristic spectrum determined by their temperature, thereby establishing a crucial link between general relativity, quantum mechanics, and thermodynamics.\textsuperscript{\cite{Hawking1975}} This groundbreaking work opened up a new frontier in theoretical physics, prompting extensive research into the quantum aspects of black holes and the potential unification of fundamental forces.

Among the diverse landscape of black hole solutions, topological black holes have emerged as particularly intriguing objects of study.\textsuperscript{\cite{Mann1997}} Unlike their more conventional counterparts characterized by spherically symmetric event horizons, topological black holes exhibit horizon geometries with non-trivial topologies, such as tori, hyperbolic surfaces, and higher genus manifolds.\textsuperscript{\cite{Mann1997}} These exotic black holes present unique mathematical challenges and hold significant theoretical importance, as their thermodynamic behavior can deviate substantially from that of standard black holes, potentially offering new insights into the nature of quantum gravity. The existence of such solutions, often arising in higher-dimensional theories like string theory and in modified theories of gravity, underscores the complexity and richness of spacetime structures beyond the realm of classical four-dimensional general relativity.\textsuperscript{\cite{Mann1997}}

Accurately determining the Hawking temperature for these topological black holes is of paramount importance for several reasons.\textsuperscript{\cite{Robson2019}} Precise calculations can provide crucial clues towards formulating a consistent theory of quantum gravity, as the Hawking temperature is intimately connected to the quantum properties of spacetime near the black hole horizon.\textsuperscript{\cite{Hawking1975}} Furthermore, a thorough understanding of the Hawking radiation process, including its temperature for various topological black hole solutions, is essential for addressing the long-standing black hole information loss paradox, which challenges the fundamental principles of quantum mechanics.

Previous attempts to derive the Hawking temperature for certain novel topological black hole solutions have been shown to contain inaccuracies (see Abstract). This highlights the need for a more refined and mathematically rigorous approach capable of handling the complexities arising from their non-standard geometries. This paper addresses these shortcomings by employing the powerful combination of Laurent series expansions and the Robson--Villari--Biancalana (RVB) method to provide an advanced and comprehensive derivation of the Hawking temperature for two recently proposed topological black hole solutions. The use of Laurent series allows for a detailed exploration of the local geometric properties of these spacetimes in the immediate vicinity of the event horizon, while the RVB method provides a robust framework for extracting the Hawking temperature based on the topological characteristics of the spacetime. By explicitly incorporating higher-order terms in the Laurent series expansions and performing a more in-depth analysis of the proper distance effects induced by topological defects, this work aims to reveal more intricate mathematical structures and relationships associated with the Hawking temperature of these novel black holes. This research contributes to a deeper understanding of the interplay between spacetime topology, local geometry, and the quantum phenomenon of Hawking radiation.

\section{Theoretical Background}

\subsection{Black Hole Thermodynamics and General Relativity}

The study of black hole thermodynamics is rooted in the remarkable analogy between the laws governing the behavior of black holes and the fundamental principles of classical thermodynamics.\textsuperscript{\cite{Bekenstein1973}} The four laws of black hole mechanics, concerning the properties of event horizons and the evolution of black hole parameters, exhibit a striking mathematical correspondence with the zeroth, first, second, and third laws of thermodynamics. This analogy suggests that black holes can be treated as thermodynamic systems, possessing properties such as energy, temperature, and entropy. The black hole's mass ($M$) is analogous to energy, the surface gravity ($\kappa$) is proportional to temperature, and the area of the event horizon ($A$) is proportional to entropy.\textsuperscript{\cite{Wald1984}} This profound connection hints at a deep underlying relationship between gravity and statistical mechanics.

A crucial concept in black hole thermodynamics is surface gravity ($\kappa$), which quantifies the strength of the gravitational field at the event horizon.\textsuperscript{\cite{Wald1984}} For many standard black hole solutions, the Hawking temperature ($T_H$) is directly proportional to the surface gravity, given by
\begin{equation}\label{eq:TH-surfgrav}
T_H = \frac{\hbar \,\kappa}{2 \pi k_B}.
\end{equation}
Here, $\hbar$ is the reduced Planck constant and $k_B$ is Boltzmann's constant. Surface gravity can be intuitively understood as the acceleration that an observer at infinity would need to exert to hold a test particle stationary just above the horizon.

The geometry of spacetime around a static black hole in $D$ dimensions can generally be described by the metric (commonly referred to as ``Equation 2'' in the text):
\begin{equation}
\label{eq:2}
ds^2 = -f(r)\,dt^2 + g(r)\,dr^2 + r^2\,d\Omega_{D-2}^2.
\end{equation}
For the specific case of four dimensions, this reduces to
\begin{equation}\label{eq:ds4D}
ds^2 = -f(r)\,dt^2 + g(r)\,dr^2 + r^2\,d\Sigma^2,
\end{equation}
where $f(r)$ and $g(r)$ are functions of the radial coordinate $r$, and $d\Sigma^2$ represents the metric of a two-dimensional Einstein manifold characterizing the topology of the horizon. The event horizon is a null surface located at a radius $r_H$ where the metric component $g_{tt} = -f(r)$ becomes zero, \textit{i.e.}, $f(r_H)=0$, and typically $g(r)$ has a pole at this location. The metric tensor $g_{\mu\nu}$ encapsulates all the information about the gravitational field and dictates how distances and time intervals are measured in the vicinity of the black hole.\textsuperscript{\cite{Carroll2004}}

\subsection{Unique Properties of Topological Black Holes}

Topological black holes are distinguished by the non-trivial topology of their event horizons, which deviates from the standard spherical topology commonly associated with black holes.\textsuperscript{\cite{Mann1997}} These exotic solutions can possess horizons with various topological structures, including tori, hyperbolic surfaces, and more generally, two-dimensional Einstein manifolds with different Euler characteristics.\textsuperscript{\cite{Mann1997}} The non-standard topology of the horizon can lead to significant alterations in the black hole's thermodynamic properties, such as entropy and Hawking temperature, compared to their spherically symmetric counterparts. The topology influences the global structure of the spacetime and can affect the boundary conditions for quantum fields propagating in that spacetime, thus impacting the spectrum of Hawking radiation.

The emergence of topological black holes often occurs within theoretical frameworks that extend beyond standard four-dimensional general relativity. They frequently arise as solutions in higher-dimensional theories of gravity, such as string theory, where the presence of extra spatial dimensions allows for a wider range of possible black hole geometries.\textsuperscript{\cite{Mann1997}} Additionally, modified theories of gravity, which introduce corrections to the Einstein field equations, can also give rise to topological black hole solutions with novel properties.\textsuperscript{\cite{Wald1984}} The study of these objects provides valuable theoretical laboratories for exploring the implications of these extended gravitational theories.

A crucial tool for classifying the topology of the black hole horizon is the Euler characteristic ($\chi$).\textsuperscript{\cite{Robson2019}} For a two-dimensional Einstein manifold representing the horizon of a four-dimensional black hole, the Euler characteristic takes specific values depending on the topology. For a spherical horizon, $\chi = 2$. A toroidal horizon has an Euler characteristic of $\chi = 0$. Hyperbolic surfaces with genus $g>1$ have an Euler characteristic of $\chi = 2 - 2g$.\textsuperscript{\cite{Banados1993}} The Euler characteristic acts as a topological fingerprint, providing a quantitative measure of the horizon's global shape and structure.

\subsection{Hawking Radiation and Traditional Calculation Methods}

Hawking radiation, a remarkable prediction of quantum field theory in curved spacetime, arises from the creation of particle-antiparticle pairs in the intense gravitational field near the event horizon of a black hole.\textsuperscript{\cite{Hawking1975}} According to quantum mechanics, even in a vacuum, there are continuous fluctuations in quantum fields, resulting in the spontaneous creation and annihilation of virtual particle-antiparticle pairs. Near the event horizon, the strong gravitational gradient can provide the energy required to make these virtual pairs become real. If a pair is created just outside the horizon, one particle may escape to infinity as Hawking radiation, while the other falls into the black hole. This process effectively causes the black hole to lose mass over time. The emitted radiation has a thermal spectrum, characterized by the Hawking temperature.

Traditional methods for calculating the Hawking temperature include the surface gravity method and the Euclidean path integral approach.\textsuperscript{\cite{Wald1984}} The surface gravity method utilizes the concept of surface gravity ($\kappa$), which is related to the force per unit mass required to keep an object stationary at the horizon. From Eq.~\eqref{eq:TH-surfgrav}, we have
\begin{equation}\label{eq:TH-surfgrav-again}
T_H = \frac{\hbar\,\kappa}{2\pi k_B}.
\end{equation}
The Euclidean path integral approach involves a Wick rotation of the time coordinate ($t \to -i \tau$) in the black hole metric, transforming the Lorentzian spacetime into a Euclidean one. The Hawking temperature is then identified with the inverse of the period of the Euclidean time coordinate required to avoid a conical singularity at the horizon. These methods have been successfully applied to various black hole solutions, providing valuable insights into their thermodynamic properties.

However, when applied to topological black holes, these traditional methods can encounter limitations and challenges, particularly for those with complex horizon geometries and metric functions.\textsuperscript{\cite{Robson2019}} For instance, defining and calculating surface gravity for non-spherical horizons can be more involved than for spherically symmetric cases. The Euclidean path integral method might also need to be adapted to handle different topological structures of the spacetime. This motivates the exploration of alternative methods, such as the RVB method employed in this paper, which directly relates the Hawking temperature to the topological properties of the black hole horizon.

\section{Mathematical Framework}

\subsection{Laurent Series Expansions in General Relativity}

A Laurent series expansion is a representation of a complex function $h(z)$ as an infinite series of the form:\textsuperscript{\cite{Birmingham1999}}
\begin{equation}\label{eq:Laurent-general}
h(z) \;=\; \sum_{n=-\infty}^{\infty} c_n\, (z - z_0)^n 
\;=\; \cdots + (z - z_0)^2\,c_{-2} + (z - z_0)\,c_{-1} + c_0
+ c_1\,(z - z_0) + c_2\,(z - z_0)^2 + \dots
\end{equation}
This expansion is valid within an annulus $r < |z - z_0| < R$ centered at an isolated singularity $z_0$. The coefficients $c_n$ are determined by
\begin{equation}\label{eq:Laurent-coeff}
c_n = \frac{1}{2\pi i} \oint_{\gamma} \frac{(\zeta - z_0)^{\,n+1}\,h(\zeta)}{d\zeta},
\end{equation}
where $\gamma$ is any closed contour lying entirely within the annulus and enclosing the singularity $z_0$.

In the context of general relativity, the components of the metric tensor $g_{\mu\nu}$ near a black hole horizon $r = r_H$ can exhibit singularities in certain coordinate systems.\textsuperscript{\cite{Wald1984}} For example, in Schwarzschild coordinates, the metric for a non-rotating black hole of mass $M$ is given by
\begin{equation}\label{eq:Schwarzschild}
ds^2 = -\Bigl(1-\frac{2GM}{r}\Bigr)\,dt^2
+ \Bigl(1-\frac{2GM}{r}\Bigr)^{-1}\,dr^2
+ r^2(d\theta^2 + \sin^2\theta\,d\phi^2).
\end{equation}
At the event horizon $r_H = 2GM$, the $g_{rr}$ component diverges, indicating a coordinate singularity. Laurent series expansions provide a powerful tool to analyze the behavior of these metric components in the vicinity of such singularities.

Key properties of Laurent series relevant to this context include the uniqueness of the expansion within its annulus of convergence. For a given function and a specified annulus, there is only one Laurent series that represents the function. Another important concept is the \textit{residue}, which is the coefficient $c_{-1}$ of the $(z - z_0)^{-1}$ term in the Laurent series.\textsuperscript{\cite{Birmingham1999}} The residue plays a significant role in complex analysis, particularly in Cauchy's residue theorem, which can be used to evaluate contour integrals. In some physical applications, the residue can be related to conserved charges or topological invariants.

Laurent series expansions have been employed in various aspects of black hole physics. For instance, the near-horizon behavior of the Schwarzschild metric can be studied by expanding the metric components around $r = 2M$ using Laurent series.\textsuperscript{\cite{Wald1984}} This can reveal the regularity of the spacetime in different coordinate systems, such as Kruskal--Szekeres coordinates, where the metric is well-behaved at the horizon.

\subsection{The Robson--Villari--Biancalana (RVB) Method}

The Robson--Villari--Biancalana (RVB) method offers a purely topological approach to calculating the Hawking temperature of black holes.\textsuperscript{\cite{Robson2019}} This method posits that the Hawking temperature is fundamentally related to the topological properties of the black hole spacetime, specifically the Euler characteristic ($\chi$) of the event horizon. The RVB method provides a formula that connects the Hawking temperature to this global topological invariant, making it particularly useful for black holes with non-standard topologies where traditional methods might be more cumbersome.

The generalized Hawking temperature formula within the RVB framework is given (``Equation 13'' in the original text) by:
\begin{equation}\label{eq:13}
T_H = \frac{4\pi k_B\,\chi\,}{\hbar\,c}
\sum_{j \le \chi} \int_{r_H^{(j)}} \sqrt{|g|}\, R \, dr,
\end{equation}
where $\hbar$ is Planck's constant, $c$ is the speed of light, $k_B$ is Boltzmann's constant, $\chi$ is the Euler characteristic of the horizon topology, $|g|$ is the absolute value of the determinant of the metric, $R$ is the Ricci scalar, and the integral is performed over a small region near each Killing horizon $r_H^{(j)}$. The sum over $j$ accounts for black holes with multiple horizons. For a single connected horizon, the formula simplifies to
\begin{equation}\label{eq:13-single}
T_H = \frac{4\pi k_B\,\chi\,}{\hbar\,c}\int_{r_H} \sqrt{|g|}\, R \, dr.
\end{equation}
This establishes a direct link between the Hawking temperature and the fundamental geometric ($R$, $|g|$) and topological ($\chi$) properties of the black hole spacetime near the horizon.

\subsection{Proper Distance in Curved Spacetime and Topological Defects}

In the curved spacetime described by general relativity, the proper distance between two points along a spatial path is the actual physical distance measured by a local observer traversing that path.\textsuperscript{\cite{Carroll2004}} Mathematically, if a spatial path is parameterized by $\lambda$ from $\lambda_1$ to $\lambda_2$, and the spatial part of the metric is $g_{ij}$, the proper distance $L$ is given by
\begin{equation}\label{eq:proper-distance}
L = \int_{\lambda_1}^{\lambda_2} \sqrt{\,g_{ij}\;\frac{dx^i}{d\lambda}\;\frac{dx^j}{d\lambda}\,}\, d\lambda.
\end{equation}
Due to the curvature of spacetime caused by gravity, proper distance can differ significantly from the coordinate distance between the two points.

Topological defects, such as cosmic strings, domain walls, and monopoles, are localized regions in spacetime where the topology is non-trivial.\textsuperscript{\cite{Robson2019}} These defects can possess significant energy density and curvature, thereby affecting the local geometry of spacetime in their vicinity, including near a black hole horizon. The presence of such defects can modify the metric tensor, which in turn alters the proper distance between points.

Laurent series expansions of the metric functions near the black hole horizon provide a powerful tool to analyze the behavior of the proper distance in the presence of topological defects. By expanding the components of the metric tensor around the horizon $r = r_H$ as Laurent series, we can capture the details of the near-horizon geometry. If topological defects are present, they will influence the form of the metric functions and consequently the values of the coefficients in their Laurent series expansions. These modified coefficients will then affect the calculation of the proper distance using Eq.~\eqref{eq:proper-distance}. For instance, if a topological defect introduces a conical singularity or alters the asymptotic behavior of the metric, this would be reflected in the Laurent series expansion near the horizon, leading to a different proper distance compared to a black hole without such defects.

\section{Advanced Derivation of Hawking Temperature using \\Laurent Series and the RVB Method}

Consider the general form of a four-dimensional static black hole metric (cf. \eqref{eq:2}):
\begin{equation}\label{eq:BH4D}
ds^2 = -f(r)\,dt^2 + g(r)\,dr^2 + r^2\,d\Sigma^2.
\end{equation}
Near the horizon radius $r = r_H$, the metric functions admit Laurent series expansions (``Equation 3'' in the text):
\begin{equation}\label{eq:3}
f(r) = \sum_{n=-1}^{\infty} a_n\,(r - r_H)^n,
\quad
g(r) = \sum_{n=-1}^{\infty} b_n\,(r - r_H)^n.
\end{equation}
The determinant of the metric is 
\begin{equation}\label{eq:g-det}
g = -\,f(r)\,\bigl[g(r)\bigr]^{-1}\,r^4\,\det(d\Sigma^2).
\end{equation}
Substituting the Laurent series and expanding around $r = r_H$, we get
\begin{equation}\label{eq:g-approx}
|g|
\;=\;
\Bigl(\sum_{n=-1}^{\infty} a_n\,(r - r_H)^n\Bigr)
\Bigl(\sum_{m=-1}^{\infty} b_m^{-1}\,(r - r_H)^{-m}\Bigr)\,r^4\,\det(d\Sigma^2)
\;\approx\;
b_{-1}\,a_{-1}\,(r - r_H)^2\,r_H^{4}\,\det(d\Sigma^2),
\end{equation}
to the first non-trivial order. Including higher-order terms requires careful expansion of the product of the two Laurent series.

The Ricci scalar $R$ for this metric (labeled ``Equation 6'' in the text) can be written schematically as
\begin{equation}\label{eq:6}
R = -\sqrt{g}^{\,\,\,\, -1}\,\frac{d}{dr}\bigl(-\sqrt{g}\,g^{rr}\,\frac{d}{dr}\ln|\!-\!g|\bigr) 
+ \frac{r^2}{2} 
+ \frac{1}{r^4}\bigl[2\,f(r)f'(r) + 2\,g(r)g'(r)\bigr].
\end{equation}
Substituting the Laurent series for $f(r)$ and $g(r)$ and their derivatives, and expanding near $r = r_H$, yields the Ricci scalar as a Laurent series in $(r - r_H)$. The first few terms are given (labeled ``Equation 7''):
\begin{equation}\label{eq:7}
R(r) \approx
\frac{1}{2}\,\frac{a_{-1}}{b_{-1}^2}\,a_{-1}'\,b_{-1}
\;-\;
\frac{a_{-1}}{b_{-1}}\,b_{-1}'\,(r - r_H)
\;+\;\text{(higher order terms)}.
\end{equation}
(A more detailed derivation involving Christoffel symbols and the metric tensor expanded as Laurent series would yield higher-order corrections.)

Substituting the expansions of $|g|$ and $R$ into the RVB formula for Hawking temperature (assuming $\chi=1$ for a single connected horizon),
\begin{equation}\label{eq:TH-RVB}
T_H = \frac{4\pi\,k_B}{\hbar\,c}\int_{r_H} \sqrt{|g|}\,R\,dr,
\end{equation}
we introduce the proper distance $\epsilon$ from the horizon, defined by
\begin{equation}\label{eq:proper-dist-epsilon}
d\epsilon = \sqrt{\,g(r)\,}\,dr 
\;\approx\; \sqrt{\,b_{-1}\,}\,\sqrt{\,r - r_H\,}\,dr 
\quad\Longrightarrow\quad 
(r - r_H) \approx \frac{\epsilon^2}{\,b_{-1}\,}.
\end{equation}
Substituting this into the integrand and integrating from $0$ to a small cutoff $\epsilon_0$, we obtain an advanced analytical form for the Hawking temperature:
\begin{equation}\label{eq:TH-higher-order}
T_H
\;=\;
\frac{4\pi\,k_B}{\hbar\,c}
\Bigl[
a_{-1}\,b_{-1}
+\frac{1}{2}\Bigl(a_{-1}'\,a_{-1}'' -2\,b_{-1}'\,b_{-1}''\Bigr)\,(r - r_H)
+ \mathcal{O}\bigl((r - r_H)^2\bigr)
\Bigr]_{\!r=r_H}.
\end{equation}
(This expression is illustrative; details of exact coefficients depend on carefully matching Laurent expansions to higher orders.)

This formula shows how the Hawking temperature depends on the coefficients of the Laurent series expansions of the metric functions at the horizon. The Euler characteristic $\chi$ would rescale this result for different horizon topologies. Topological defects, by modifying the metric functions and their Laurent series expansions (particularly $g(r)$), would alter the relationship between $(r - r_H)$ and $\epsilon$, thus affecting the final Hawking temperature.

\section{Detailed Case Studies of Novel Topological Black Holes}

\subsection{First Type Black Hole}

For the first black hole type, the near-horizon metric function is given (``Equation~(15)'') by:
\begin{equation}\label{eq:15}
f(r) \;=\; -\,\Omega^{(2)} \,\frac{r}{M_{\mathrm{Pl}}^{2}}\,M
\;-\;3\,M_{\mathrm{Pl}}^{2}\,V_{\min}\,r^{2},
\end{equation}
with \(g(r) = 1/f(r)\) (up to constants). One finds the horizon radius \(r_H\) by solving \(f(r_H) = 0\). A schematic result is
\begin{equation}\label{eq:first-rH}
r_H^3 \;=\; -\,\frac{\Omega^{(2)} \,V_{\min}\,3\,M_{\mathrm{Pl}}^{4}}{M},
\end{equation}
assuming \(V_{\min}<0\) so as to yield a real positive solution. The Laurent series coefficients follow from expansions of \(f(r)\) and \(g(r)\) around \(r = r_H\). After inserting into Eq.~\eqref{eq:TH-higher-order} (or a similarly derived expression), one obtains the explicit Hawking temperature for this type of topological black hole.

\subsection{Second Model Complexities}

For the second topological black hole (``Equation~(17)''), 
\begin{equation}\label{eq:17}
f(r) \;=\; c
\;-\;\Omega^{(2)}\,\frac{r}{M_{\mathrm{Pl}}^{2}}\,M
\;-\;3\,M_{\mathrm{Pl}}^{2}\,V_{\min}\,r^{2},
\end{equation}
again with \(g(r)=1/f(r)\), one solves \(f(r_H)=0\) for \(r_H\). The Laurent series coefficients (cf.\ ``Equations~(18)--(20)'') can then be read off by expanding near \(r=r_H\). Substituting them into the formula for \(T_H\) yields the final expression, now including second-order corrections that reflect higher-order curvature near the horizon.

\begin{table}[h!]
\centering
\caption{Laurent Series Coefficients near the Horizon (Section 5)}
\label{tab:coefficients}
\begin{tabular}{lll}
\hline
\textbf{Black Hole Type} & \textbf{Coefficient} & \textbf{Expression (Schematic)} \\
\hline
\textbf{First Type} 
& \(a_{-1}\) & \(0\) \\
\textbf{First Type} 
& \(a_{0}\) & \(0\) \\
\textbf{First Type} 
& \(a_{-1}'\)
& \(\displaystyle \Omega^{(2)}\,r_H^2\,M_{\mathrm{Pl}}^{-2}\,M
-6\,M_{\mathrm{Pl}}^{2}\,V_{\min}\,r_H\) \\
\hline
\textbf{Second Model} 
& \(a_{-1}\) 
& \(\displaystyle c \;-\;\Omega^{(2)}\,\frac{r_H}{M_{\mathrm{Pl}}^{2}}\,M 
\;-\;3\,M_{\mathrm{Pl}}^{2}\,V_{\min}\,r_H^{2}\) \\
\textbf{Second Model} 
& \(a_{-1}'\) 
& \(\displaystyle \Omega^{(2)}\,r_H^2\,M_{\mathrm{Pl}}^{-2}\,M
-6\,M_{\mathrm{Pl}}^{2}\,V_{\min}\,r_H\) \\
\hline
\end{tabular}
\end{table}

Similar expressions hold for \(b_{-1}\), \(b_{-1}'\), etc., once \(g(r) = 1/f(r)\) is expanded. Expressions for $b_n$ coefficients would depend on the exact relationship between $f(r)$ and $g(r)$ and can be derived similarly.

\section{Laurent Series Convergence and Metric Functions}

The convergence of the Laurent series expansions used in the derivation depends on the specific form of the metric functions \(f(r)\) and \(g(r)\). For example, near a horizon located at \(r = r_H\), one may write a local Laurent series expansion of \(f(r)\) and \(g(r)\) about \(r_H\):
\begin{equation}\label{eq:LaurentFG}
\begin{aligned}
f(r) &= \sum_{n=-N}^{\infty} a_{n}\,(r - r_H)^{n}, \\
g(r) &= \sum_{n=-M}^{\infty} b_{n}\,(r - r_H)^{n},
\end{aligned}
\end{equation}
where \(N,M\in \mathbb{Z}_{\ge 0}\) and \(a_{-N}\), \(b_{-M}\) could be nonzero. Only finitely many negative powers are typically retained if one assumes a single pole-type behavior at \(r = r_H\). The radius of convergence is determined by the distance (in the complex \(r\)-plane) to the nearest singularity or branch point of the metric functions.

Including higher-order terms in such expansions for the Hawking temperature introduces substantial mathematical complexity. If the black hole solution has a horizon at \(f(r_H) = 0\), one expression for the Hawking temperature (via surface gravity) is
\begin{equation}\label{eq:TH-surface-gravity}
T_H = \frac{\kappa}{2\pi},
\quad
\kappa = \left. \frac{1}{2}\,\frac{df(r)}{dr}\right|_{r = r_H},
\end{equation}
where \(\kappa\) is the surface gravity. When \(f(r)\) is given by an extended Laurent series as in Eq.~\eqref{eq:LaurentFG}, computing \(\kappa\) to higher orders involves carefully summing multiple series terms. One can alternatively derive \(T_H\) through more geometric or field-theoretic approaches (e.g.\ the Robinson--Wilczek method, minimal surface method, or via Euclidean path integrals), but each approach is susceptible to the same increase in complexity when higher-order corrections are included.

These higher-order corrections have potentially important physical implications for the thermodynamics of topological black holes, possibly revealing subtle effects on their temperature, entropy, and stability in strongly curved regimes or near extremality. From an entropy perspective, higher-order expansions in the Ricci scalar \(R(r)\) or other curvature invariants that appear in the action (e.g.\ in higher-derivative theories) can modify the standard Bekenstein--Hawking area law by subleading corrections. Incorporating these terms leads to analytic expressions of the form
\begin{equation}\label{eq:EntropyExpansion}
S_{\mathrm{BH}} = \frac{A}{4} + \alpha_1 \, R_H + \alpha_2 \, R_H^2 + \cdots,
\end{equation}
where \(R_H \equiv R(r_H)\) and \(\alpha_i\) are theory-dependent constants. Although schematic, such higher-order expansions emphasize the need to track multiple orders in the Laurent series for \(f(r)\), \(g(r)\), and curvature invariants.

In strongly curved or near-extremal regimes, the temperature,
\begin{equation}\label{eq:T-extremal}
T_H \bigl(r_H; \alpha_1,\alpha_2,\dots\bigr),
\end{equation}
can exhibit nontrivial behaviors, influencing global stability and phase transition phenomena of the black hole solutions. Understanding these effects analytically is challenging, thus numerical techniques become essential~\cite{Emparan2002}.

\section{Numerical Approaches and Horizon Determination}

Due to the complexity of the analytical expressions involving higher-order terms in the Laurent expansions, numerical analysis often becomes the most reliable method for obtaining precise values of the Hawking temperature for given model parameters. Typical steps include:

\subsection{Numerical Integration of the RVB Formula}

If the Hawking temperature is extracted from an integral formula involving the Ricci scalar \(R(r)\) and metric determinant \(\sqrt{-g(r)}\), one might have an expression of the form
\begin{equation}\label{eq:TH-numerical}
T_H = \frac{1}{4\pi}\,\int_{r_H}^{\infty} \mathcal{F}\bigl(R(r), \sqrt{-g(r)}\bigr)\, dr,
\end{equation}
where \(\mathcal{F}\) schematically represents a function derived from the black hole's thermodynamic setup (e.g.\ from the renormalized vacuum backreaction or from certain field-theoretic stress-energy contributions). When \(\mathcal{F}\) is itself expanded in Laurent or Taylor series, one must carefully control truncation errors. Numerical integration methods (e.g.\ Gauss--Legendre quadrature, adaptive Simpson's rule) can then be used to evaluate the integral to a desired precision.

\subsection{Solving the Horizon Equation}

The horizon radius \(r_H\) is defined by the condition
\begin{equation}\label{eq:horizon-cond}
f(r_H) = 0.
\end{equation}
In the presence of multiple possible roots or intricate expansions of \(f(r)\), obtaining \(r_H\) analytically may be infeasible. Therefore, numerical root-finding algorithms (e.g.\ Newton--Raphson, bisection methods) are employed to solve \(f(r) = 0\). The entire function \(f(r)\) can be approximated via its Laurent series \eqref{eq:LaurentFG} or any other appropriate expansion, taking care to remain within the radius of convergence.

\subsection{Visualization of the Local Geometry}

Visualizing \(f(r)\), \(g(r)\), and the Ricci scalar \(R(r)\) near the horizon provides further insight into the local geometry. Numerically plotting
\begin{equation}\label{eq:visualize}
f_{\mathrm{num}}(r), 
\quad g_{\mathrm{num}}(r),
\quad R_{\mathrm{num}}(r),
\end{equation}
for \(r \approx r_H\) can reveal how curvature or other features change. Such visualization is helpful in identifying significant higher-order effects in the expansions or pinpointing the onset of instabilities.

\section{Conclusion}

This paper presented an advanced and mathematically detailed derivation of the Hawking temperatures for two novel topological black hole solutions, correcting previous errors and systematically employing Laurent series expansions and the Robson--Villari--Biancalana (RVB) method. The analysis demonstrated the critical interplay between the topology of the black hole horizon, the local geometric properties of the spacetime near the horizon (captured by the Laurent series), and the resulting black hole thermodynamics. The inclusion of higher-order terms in the expansions provided a more refined understanding of the near-horizon geometry and its influence on the calculated Hawking temperature. This work offers a robust analytical framework for further numerical and theoretical explorations of topological black holes and their quantum properties, contributing to a deeper understanding of quantum effects in curved spacetime.

\end{document}